\documentclass[10pt]{mpi08} \usepackage{graphicx} 
\usepackage{cite,./mcite}  
\usepackage{graphicx}
\usepackage{amssymb,amsmath}
\usepackage{units}

\begin{document} \title{(Multiple) Hard Parton Interactions in Heavy-Ion Collisions} 
\author{Klaus Reygers}
\institute{Physikalisches Institut, Universit\"{a}t Heidelberg, 
  Philosophenweg 12,\\69120 Heidelberg, Germany}
\maketitle

\begin{abstract} 
  Multiple hard interactions of partons in the same $p+p(\bar{p})$
  collision are a useful concept in the description of these
  collisions at collider energies. In particular, they play a crucial
  role for the understanding of the background (the so-called
  underlying event) in the reconstruction of jets.  In nucleus-nucleus
  collisions multiple hard parton interactions and the corresponding
  production of mini-jets are expected to contribute significantly to
  the total particle multiplicity. In this article a brief overview of
  results on particle production at high-$p_\mathrm{T}$ in
  proton-proton and nucleus-nucleus at RHIC will be given. Moreover,
  the observed centrality dependence of the charged particle
  multiplicity in Au+Au collisions will be discussed in light of
  multiple partonic interactions.
\end{abstract}

\section{Introduction} 
In a $p+p(\bar{p})$ collision the location of a hard parton-parton
scattering in which a parton with transverse momentum $p_\mathrm{T}
\gtrsim \unit[2]{GeV}/c$ is produced is well defined ($\Delta r \sim
1/p_\mathrm{T} \lesssim \unit[0.1]{fm}$ in the plane transverse to the
beam axis) and much smaller than the radius of proton ($r \approx
\unit[0.8]{fm}$). Thus, it is expected that multiple hard parton
scatterings can contribute incoherently to the total hard scattering
cross section \cite{Sjostrand:1987su,Sjostrand:2004pf}. When going
from p+p to nucleus-nucleus (A+A) collisions and neglecting nuclear
effects the increase in the number of hard scatterings is given by the
nuclear geometry expressed via the nuclear overlap function
$T_\mathrm{AB}$ \cite{Miller:2007ri}. For a given range of the impact
parameter $b$ of the A+A collisions the yield of produced partons with
a transverse momentum $p_\mathrm{T}$ can thus be calculated from the
corresponding cross section in p+p collisions according to
\begin{equation}
  \frac{1}{N_\mathrm{inel}^\mathrm{A+A}} \left.
    \frac{\mathrm{d}N}{\mathrm{d}p_\mathrm{T}} \right|_\mathrm{A+A} =
  \frac{\int \mathrm{d}^2b \,T_\mathrm{AB}(b)} {\int \mathrm{d}^2b
    \left(1-\exp \left(-T_\mathrm{AB} \cdot
        \sigma_\mathrm{inel}^\mathrm{NN}\right)\right)} \cdot \left.
    \frac{\mathrm{d}\sigma}{\mathrm{d}p_\mathrm{T}}
  \right|_\mathrm{p+p} 
\label{eq:tab_scaling}
\end{equation}
where $N_\mathrm{inel}^\mathrm{A+A}$ denotes the total number of
inelastic A+A collisions and $\sigma_\mathrm{inel}^\mathrm{NN}$ the
inelastic nucleon-nucleon cross section.  This corresponds to a
scaling of the yield of produced high-$p_\mathrm{T}$ partons (and
hence also of the yield of hadrons at high $p_\mathrm{T}$) with the
number of binary nucleon-nucleon collisions ($N_\mathrm{coll}$). On
the other hand, the yield of particles at low $p_\mathrm{T} \lesssim
\unit[1]{GeV}/c$ is expected to scale with the number
$N_\mathrm{part}$ of nucleons that suffered at least one inelastic
nucleon-nucleon collision. Based on this separation of soft and hard
processes the centrality dependence of the charged particle
multiplicity in nucleus-nucleus collisions can be predicted.

\section{Hard Scattering at RHIC}
In this article the focus is on the study of hard scattering in p+p
and A+A collisions at RHIC by measuring particle yields at high
transverse momentum. Further methods are the statistical analysis of
2-particle angular correlations and full jet reconstruction on an
event-by-event basis \cite{Tannenbaum:2007sy, Salur:2009vz}. The
latter method is challenging in heavy-ion collisions since, {\it
  e.g.}, in a central Au+Au collision with a transverse energy of
$\mathrm{d}E_\mathrm{T}/\mathrm{d}\eta \approx \unit[500]{GeV}$ at
midrapidity the background energy from the underlying event in a cone
with a radius $R = \sqrt{(\Delta \phi)^2 + (\Delta \eta)^2 } = 0.7$ is
$ E_\mathrm{T}^\mathrm{background} \approx \unit[120]{GeV}$.  For a
general overview of result from the four RHIC experiments see
\cite{Arsene:2004fa,Back:2004je,Adcox:2004mh,Adams:2005dq}.

\begin{figure}[t]
  \includegraphics[width=\textwidth]{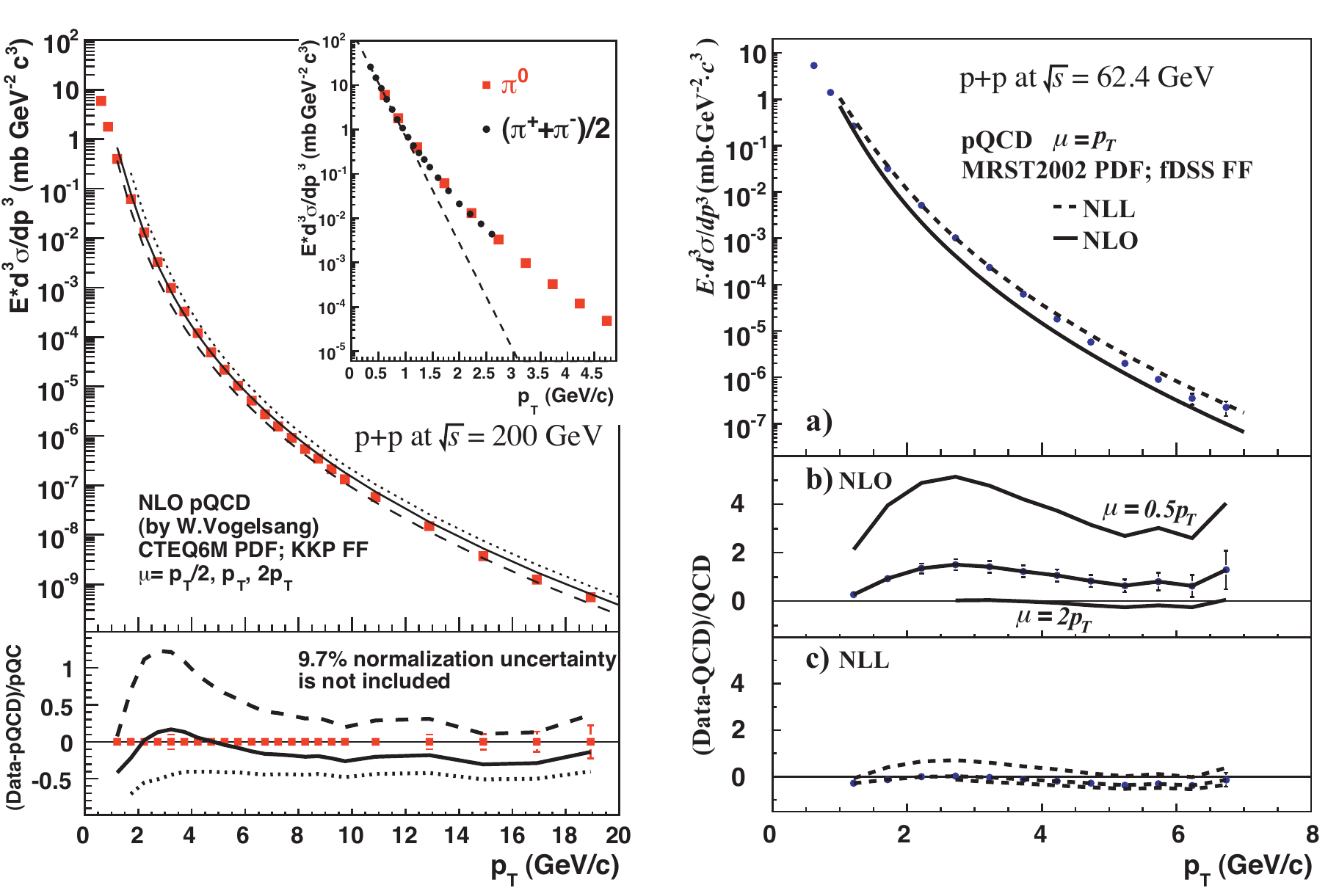}
  \caption{Invariant cross sections for the reaction $p+p \rightarrow
    \pi^0 + X$ at $\sqrt{s} = \unit[200]{GeV}$ (left panel) and
    $\sqrt{s} = \unit[62.4]{GeV}$ (right panel) as measured by the
    PHENIX experiment at RHIC \protect \cite{Adare:2007dg,
      Adare:2008qb}. The data are compared to next-to-leading-order
    (NLO) perturbative QCD calculations performed with equal
    factorization ($\mu_\mathrm{F}$), renormalization
    ($\mu_\mathrm{R}$), and fragmentation ($\mu_\mathrm{F'}$)
    scales. The theoretical uncertainties were estimated by choosing
    $\mu = \mu_\mathrm{F} = \mu_\mathrm{R} = \mu_\mathrm{F'} =
    p_\mathrm{T}, 0.5 p_\mathrm{T}, 2 p_\mathrm{T}$, respectively.}
\label{fig:pi0_spec_pp}
\end{figure}
Deviations from point-like scaling of hard processes in
nucleus-nucleus collisions described by Eq.~\ref{eq:tab_scaling} can
be quantified with the nuclear modification factor
\begin{equation}
  R_\mathrm{AA} = \frac{1/N_\mathrm{inel}^\mathrm{A+A} 
    \left. \mathrm{d}N/\mathrm{d}p_\mathrm{T}\right|_\mathrm{A+A}}
  {\langle T_\mathrm{AB} \rangle \cdot \left. \mathrm{d}\sigma/
      \mathrm{d}p_\mathrm{T}\right|_\mathrm{p+p}}
  = \frac{1/N_\mathrm{inel}^\mathrm{A+A} 
    \left. \mathrm{d}N/\mathrm{d}p_\mathrm{T}\right|_\mathrm{A+A}}
  {\langle N_\mathrm{coll} \rangle \cdot 1/N_\mathrm{inel}^\mathrm{p+p} 
    \left. \mathrm{d}N/\mathrm{d}p_\mathrm{T}\right|_\mathrm{p+p}} \:.
\label{eq:raa}
\end{equation}
Neutral pion $p_\mathrm{T}$ spectra in p+p collisions at $\sqrt{s} =
\unit[200]{GeV}$ and $\unit[62.4]{GeV}$ used in the denominator of
Eq.~\ref{eq:raa} are shown in
Fig.~\ref{fig:pi0_spec_pp}. Next-to-leading-order perturbative QCD
calculations describe the data down to $p_\mathrm{T} \approx
\unit[1]{GeV}/c$ at both energies.

\begin{figure}[t] 
  \includegraphics[width=\textwidth]{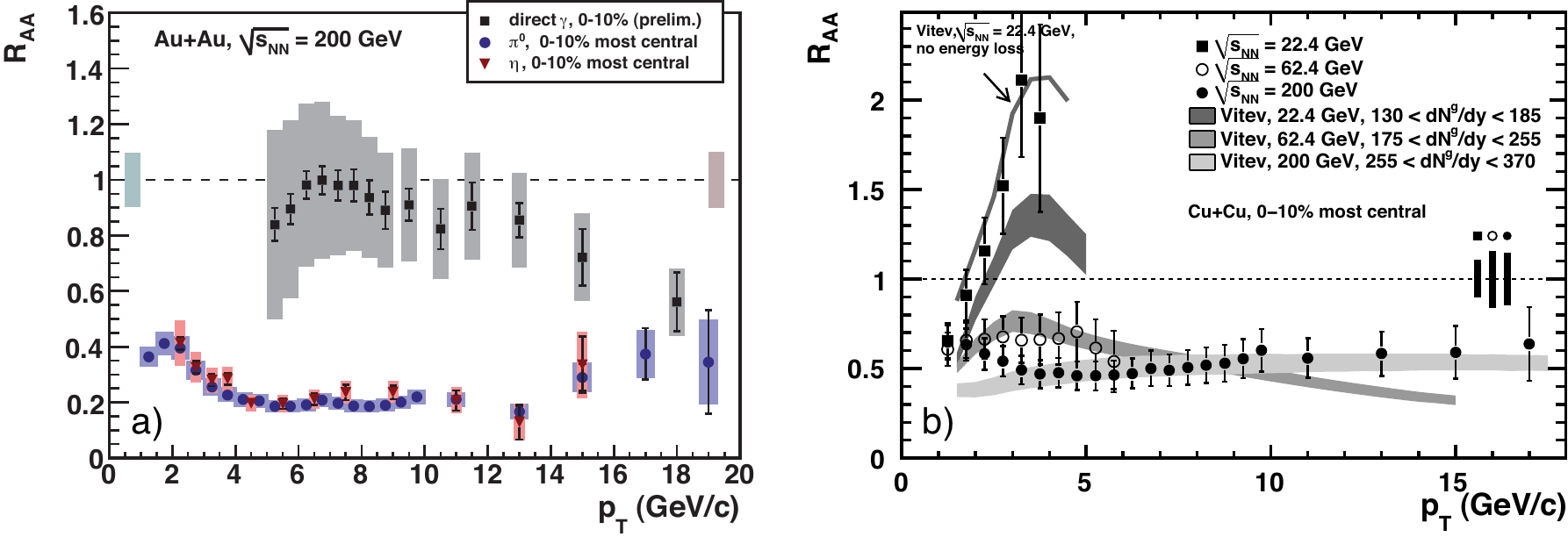}
  \caption{a) $R_\mathrm{AA}$ for $\pi^0$'s, $\eta$'s, and direct
    photons in central Au+Au collisions at $\sqrt{s_{NN}} =
    \unit[200]{GeV}$ \protect \cite{Reygers:2008pq}. b) Energy
    ($\sqrt{s_{NN}}$) dependence of $R_\mathrm{AA}$ for $\pi^0$'s in
    central Cu+Cu collisions at $\sqrt{s_{NN}} = 22.4, 62.4$ and
    $\unit[200]{GeV}/c$ \protect \cite{Adare:2008cx}.}
  \label{fig:raa} 
\end{figure}

In Au+Au collisions at $\sqrt{s_{NN}} = \unit[200]{GeV}$ a dramatic
deviation of $\pi^0$ and $\eta$ yields at high $p_\mathrm{T}$ from
point-like scaling is observed. In the sample of the 10\% most central
Au+Au collisions the yields are suppressed by a factor of $4-5$
(Fig.~\ref{fig:raa}a). On the other hand, direct photons, measured on
a statistical basis by subtracting background photons from hadron
decays like $\pi^0 \rightarrow \gamma \gamma$ or $\eta \rightarrow
\gamma \gamma$ from the $p_\mathrm{T}$ spectrum of all measured
photons, are not suppressed for $p_\mathrm{T} \lesssim
\unit[12]{GeV}/c$. Thus, one can conclude that the hadron suppression
is caused by the presence of the created hot and dense medium and is
not related to properties of cold nuclear matter.

In order to search for the onset of the high-$p_\mathrm{T}$ hadron
suppression Cu+Cu collisions at three different energies
($\sqrt{s_{NN}} = \unit[22.4, 62.4, \mathrm{and}\,200]{GeV}$) were
studied by the PHENIX experiment \cite{Adare:2008cx}. In central Cu+Cu
collisions at $\sqrt{s_{NN}} = \unit[200]{GeV}$ neutral pions at high
$p_\mathrm{T}$ are suppressed by a factor $\sim 2$
(Fig.~\ref{fig:raa}b). A similar suppression is observed at
$\sqrt{s_{NN}} = \unit[62.4]{GeV}$. However, at $\sqrt{s_{NN}} =
\unit[22.4]{GeV}$ an enhancement ($R_\mathrm{AA} > 1$) is found which
can be explained by a broadening of the transverse momentum component
of the partons in the cold nuclear medium (nuclear-$k_\mathrm{T}$ or
{\it Cronin} enhancement). The upshot is that in Cu+Cu collisions the
suppression of high-$p_\mathrm{T}$ pions sets in between
$\sqrt{s_{NN}} \approx \unit[20 - 60]{GeV}$.  In very central
collisions of heavier nuclei (Pb ions) the WA98 experiment at the CERN
SPS found a suppression of neutral pions with $p_\mathrm{T} >
\unit[2]{GeV}/c$ already at $\sqrt{s_\mathrm{NN}} = \unit[17.3]{GeV}$
\cite{Aggarwal:2007gw}.

The most likely explanation for the suppression of hadrons at high
$p_\mathrm{T}$ is energy loss of partons from hard scatterings in the
medium of high color-charge density produced nucleus-nucleus
collisions ({\it jet-quenching}) \cite{Baier:2000mf,Kovner:2003zj}. In
this picture the absolute value of the nuclear modification factor
contains information about properties of the medium such as the
initial gluon density $\mathrm{d}N^\mathrm{g}/\mathrm{d}y$. The parton
energy loss calculation shown in Fig.~\ref{fig:raa}b reproduces the
suppression in central Cu+Cu collisions at $\sqrt{s_{NN}} =
\unit[200]{GeV}$ for $255 < \mathrm{d}N^\mathrm{g}/\mathrm{d}y < 370$,
whereas the suppression in Au+Au at $\sqrt{s_{NN}} = \unit[200]{GeV}$
requires a gluon density on the order of $1250 <
\mathrm{d}N^\mathrm{g}/\mathrm{d}y < 1670$ \cite{Adare:2008cg}.

Direct photons are not expected to be suppressed in A+A collisions
since they interact only electro-magnetically with the medium and thus
have a much longer mean free path length. However, preliminary data
from the PHENIX experiment indicate a suppression in central Au+Au
collisions at $\sqrt{s_{NN}} = \unit[200]{GeV}$ also for direct
photons with $p_\mathrm{T} \gtrsim \unit[12]{GeV}$
(Fig.~\ref{fig:raa}a). This suppression can partly be explained by the
different quark content of the proton and the neutron (isospin effect)
which is not taken into account in the definition of $R_\mathrm{AA}$
\cite{Arleo:2006xb}. A further contribution might come from the
suppression of direct photons which are not produced in initial parton
scatterings but in the fragmentation of quark and gluon jets
(fragmentation photons) \cite{Arleo:2006xb}.

\begin{figure}[t]
  \includegraphics[width=\textwidth]{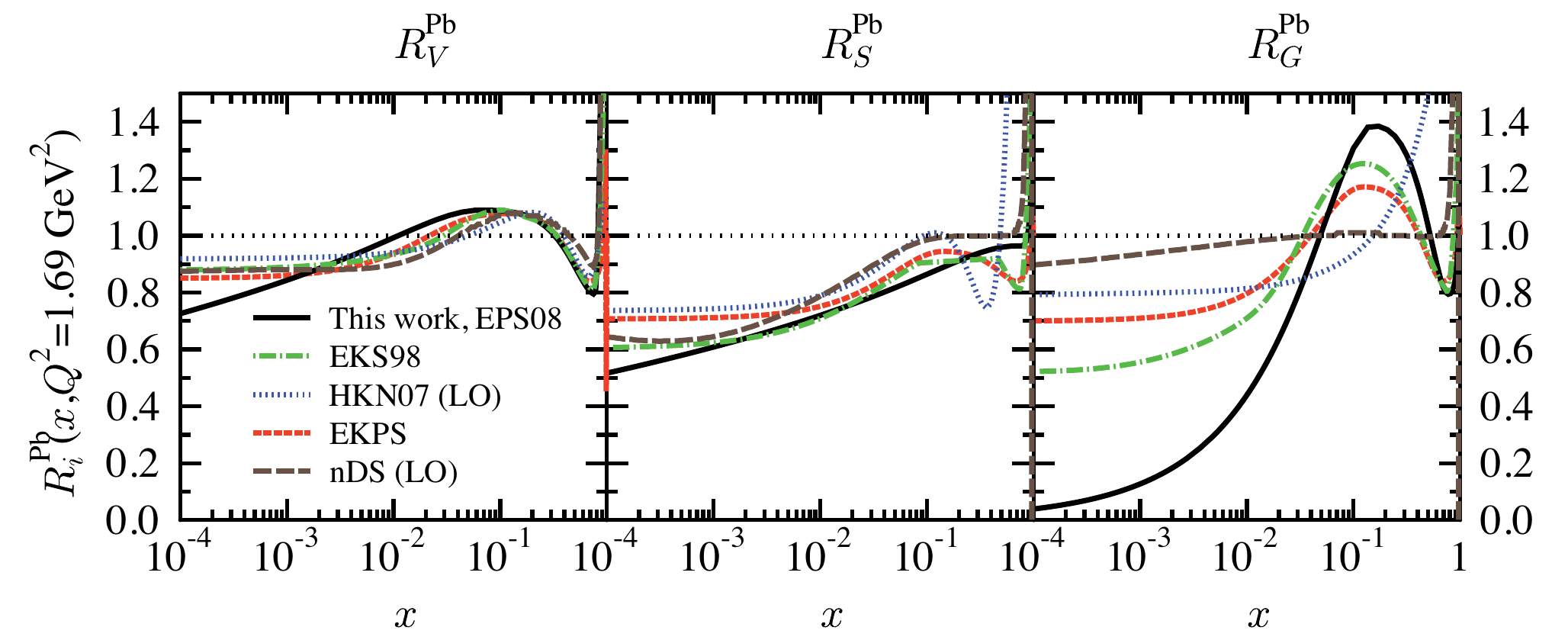}
  \caption{Different results from leading-order (LO) QCD analyses for
    the ratio of the parton distribution in the lead nucleus and in the
    proton for valence quarks (left panel), sea quark (middle panel),
    and gluons (right panel) \protect\cite{Eskola:2008ca}.}
\label{fig:npdfs}
\end{figure}

The modification of the parton distribution functions (PDF's) in the
nucleus with respect to the proton PDF's are also not taken into
account in the nuclear modification factor $R_\mathrm{AA}$. Roughly
speaking, features of nuclear PDF's as compared to proton PDF's are a
reduced parton density for $x \lesssim 0.1$ (shadowing), an
enhancement for $0.1 \lesssim x \lesssim 0.3$ (anti-shadowing)
followed again by a suppression for $0.3 \lesssim x \lesssim 0.7$
(EMC-effect) \cite{Piller:1999wx}. For $x \rightarrow 1$ the parton
densities are enhanced due to the Fermi motion of the nucleons inside
the nucleus. In Fig.~\ref{fig:npdfs} different parameterizations of
the ratio $R(x,Q^2) = f_i^\mathrm{A}(x,Q^2)/f_i^\mathrm{p}(x,Q^2)$ of
the parton distribution for a lead nucleus and for the proton are
shown for valence quarks, sea quarks, and
gluons\cite{Eskola:2008ca}. It is obvious from this comparison that
the gluon distribution in the lead nucleus is not well constrained by
lepton-nucleus deep inelastic scattering data at low $x$ ($x \lesssim
10^{-2}$).  This leads to a large uncertainty of the gluon PDF as
determined in a systematic error analysis \cite{Eskola:2009uj}.

The gluon distribution is of special interest for the understanding of
direct-photon production since quark-gluon Compton scattering $q+g
\rightarrow q+\gamma$ significantly contributes to the total
direct-photon yield. In Fig.~\ref{fig:raa}a $p_\mathrm{T} \approx
\unit[10]{GeV}/c$ where $R_\mathrm{AA}^{\mathrm{direct}\,\gamma}
\approx 1$ and $p_\mathrm{T} \approx \unit[20]{GeV}/c$ where
$R_\mathrm{AA}^{\mathrm{direct} \,\gamma} \approx 0.6$ roughly
correspond to $x \approx 0.1$ and $x \approx 0.2$, respectively,
according to $x \approx 2 p_\mathrm{T}/\sqrt{s}$. From the ratio
$R_\mathrm{G}^\mathrm{Pb}$ in this $x$ range (Fig.~\ref{fig:npdfs})
there is no indication that the suppression of direct photons at high
$p_\mathrm{T}$ in central Au+Au collisions is related to the gluon
distribution in heavy nuclei. This is in line with the calculation
presented in \cite{Arleo:2006xb}.

\section{Charged Particle Multiplicity: Hard and Soft Component}
Multiple hard partonic interaction in $p+p (\bar{p})$ collisions
explain many observed features of these collisions including the rise
of the total inelastic $p+p(\bar{p})$ cross section with $\sqrt{s}$,
the increase of $\langle p_\mathrm{T} \rangle$ with the charged
particle multiplicity $N_\mathrm{ch}$, the increase of $\langle
p_\mathrm{T} \rangle$ with $\sqrt{s}$, the increase of
$\mathrm{d}N_\mathrm{ch}/\mathrm{d}\eta$ with $\sqrt{s}$, and the
violation of KNO scaling at large $\sqrt{s}$. In such mini-jet models
a $p+p(\bar{p})$ collision is classified either as a purely soft
collision or a collision with one or more hard parton interactions
depending on a cut-off transverse momentum $p_\mathrm{T,min}$ (see
e.g.  \cite{Li:2001xa}). The cross section $\sigma_\mathrm{soft}$ for
a soft interaction is considered as a non-calculable parameter. The
energy dependence of the charged particle multiplicity in
$p+p(\bar{p})$ collisions can then be described by
\begin{equation}
\label{eq:nch_pp}
\left. \frac{\mathrm{d}N_{ch}}{\mathrm{d}\eta} \right|_{p+p} = 
\langle n_\mathrm{soft} \rangle +
\langle n_\mathrm{hard} \rangle \cdot 
\frac{\sigma_\mathrm{jet}(\sqrt{s})}{\sigma_\mathrm{inel}(\sqrt{s})}
\;.
\end{equation}
This can be extrapolated to nucleus-nucleus collisions by assuming
that the soft component scales with the number of participating
nucleons $N_\mathrm{part}$ whereas the mini-jet component scales with
the number of nucleon-nucleon collisions $N_\mathrm{coll}$:
\begin{equation}
\label{eq:nch_aa}
\left. \frac{\mathrm{d}N_{ch}}{\mathrm{d}\eta} \right|_{A+A} = 
\frac{1}{2} \langle N_\mathrm{part} \rangle \cdot 
\langle n_\mathrm{soft} \rangle +
\langle N_\mathrm{coll} \rangle \cdot
\langle n_\mathrm{hard} \rangle \cdot 
\frac{\sigma_\mathrm{jet}(\sqrt{s})}{\sigma_\mathrm{inel}(\sqrt{s})}
\;.
\end{equation}
Here $\langle n_\mathrm{soft} \rangle$ and $\langle n_\mathrm{hard}
\rangle$ are fixed parameters determined from $p+p(\bar{p})$
collisions.

\begin{figure}[t]
  \includegraphics[width=\textwidth]{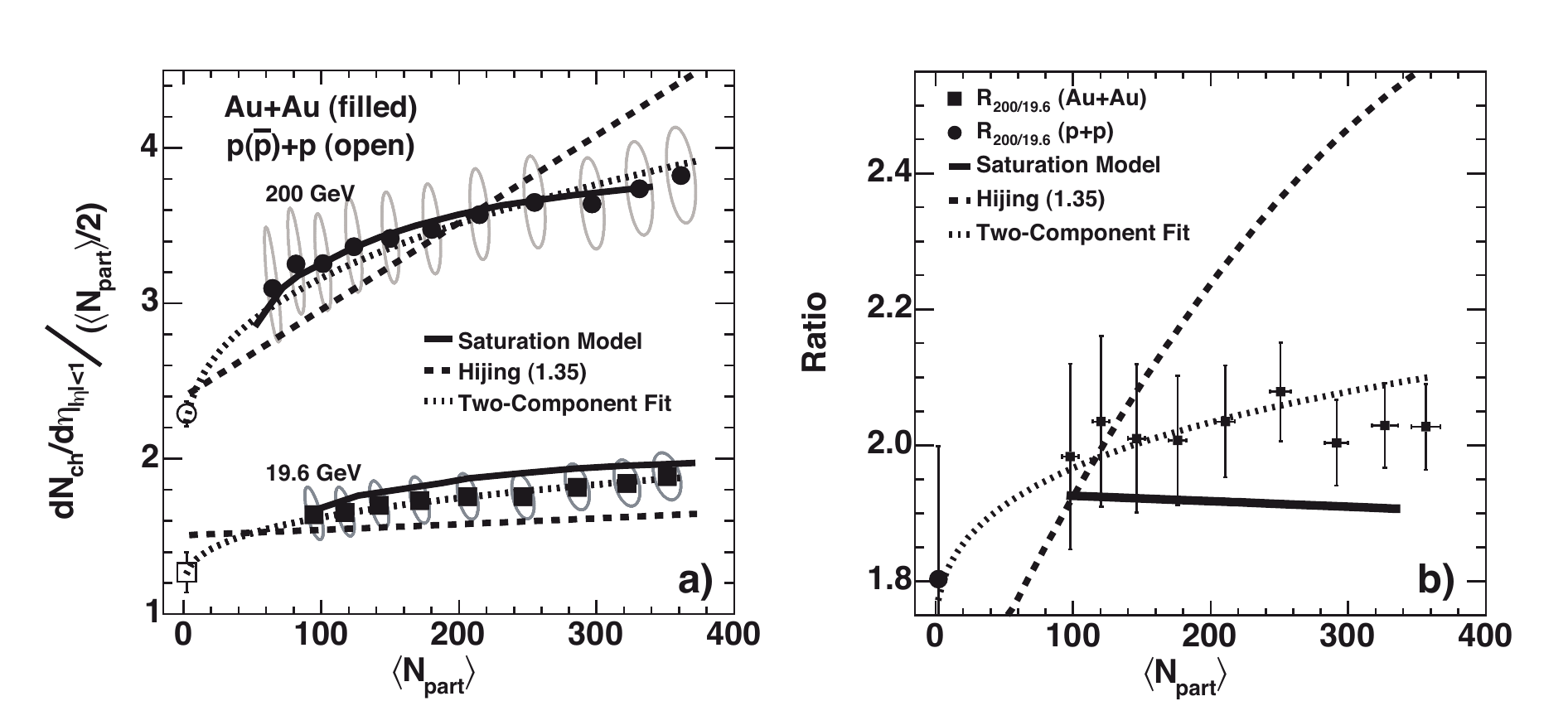}
  \caption{a) Centrality dependence of the charged particle
    multiplicity in Au+Au collisions at $\sqrt{s_\mathrm{NN}} =
    \unit[19.4]{GeV}$ and $\unit[200]{GeV}$ measured by the Phobos
    experiment \protect\cite{Back:2004je}. b) Ratio of the two data
    sets of Figure a) \protect\cite{Back:2004je}.}
  \label{fig:mult} 
\end{figure}
The centrality dependence of the charged particle multiplicity
measured in Au+Au collisions at $\sqrt{s_\mathrm{NN}} =
\unit[19.4]{GeV}$ and $\unit[200]{GeV}$ is shown in
Fig.~\ref{fig:mult}a.  Interestingly, the relative increase of the
multiplicity per participant from $\langle N_\mathrm{part} \rangle
\approx 100$ to $\langle N_\mathrm{part} \rangle \approx 350$ is
identical for the two energies. This can be described within the
experimental uncertainties with a saturation model
\cite{Kharzeev:2001gp} (Fig.~\ref{fig:mult}, solid line) and a
two-component fit which extrapolates from p+p to A+A as in
Eq.~\ref{eq:nch_aa} but leaves the relative fraction of the soft and
the hard component in p+p (Eq.~\ref{eq:nch_pp}) as a free parameter
\cite{Kharzeev:2000ph} (Fig.~\ref{fig:mult}, dotted line). However,
this behavior cannot be reproduced with the two-component mini-jet
model implemented in the Monte Carlo event generator Hijing~1.35
(Fig.~\ref{fig:mult}, dashed line). This does not necessarily mean
that the two-component picture is not valid in nucleus-nucleus
collisions as pointed out in \cite{Li:2001xa}. With the two-component
mini-jet model of ref. \cite{Li:2001xa} the experimentally observed
centrality dependence can be reproduced if a strong shadowing of the
gluon distribution in the gold nucleus is assumed. However, the used
gluon distribution deviates from the parameterizations in
Fig.~\ref{fig:npdfs} and it is stated in \cite{Li:2001xa} that with a
gluon distribution that exhibits a strong anti-shadowing as the
distributions in Fig.~\ref{fig:npdfs} the data cannot be reproduced.
Thus, the question whether the two-component mini-jet picture is a
useful concept in nucleus-nucleus collisions hinges on the knowledge
about the gluon PDF and can only be answered if the uncertainties of
the gluon distribution in nuclei can be significantly reduced.

\section{Summary}
The interest in hard scattering of partons in nucleus-nucleus
collisions is twofold: First, QCD predictions for the energy loss of
highly-energetic partons in a medium of high color-charge density can
be tested experimentally. Second, the observed hadron suppression in
conjunction with parton energy loss models renders the possibility to
characterize the medium created in ultra-relativistic nucleus-nucleus
collisions.  The assumption that indeed the created medium causes the
suppression was confirmed by the observation that direct photons at
high $p_\mathrm{T}$ which result from hard parton-parton scatterings
are not suppressed (at least for $p_\mathrm{T} \lesssim
\unit[12]{GeV}/c$ in Au+Au collisions at $\sqrt{s_{NN}} =
\unit[200]{GeV}$). It remains to be understood how the apparent
suppression of direct photons with $p_\mathrm{T} \gtrsim
\unit[12]{GeV}/c$ fits into this picture.  It was argued that it is
unlikely that this direct-photon suppression is related to the gluon
distribution function in the gold nucleus.

A natural extension of the successful concept of multiple partonic
interactions in $p+p(\bar{p})$ collisions to nucleus-nucleus
collisions is the two-component mini-jet model for the centrality
($N_\mathrm{part}$) dependence of the charged particle multiplicity.
As shown in \cite{Li:2001xa} such a model can indeed describe the
experimental data, but only if a relatively strong suppression of the
gluon distribution in a gold nucleus is assumed. The gluon
distribution in this model appears to be only barely consistent with
recent parameterizations such as EPS09LO \cite{Eskola:2009uj} so that
it remains to be seen whether the two-component mini-jet model is a
useful concept in nucleus-nucleus collisions.

\begin{footnotesize}
\bibliographystyle{mpi08} 
{\raggedright
\bibliography{mpi08}
}
\end{footnotesize}
\end{document}